\documentclass[a4paper,12pt]{article}

\usepackage{amsmath}
\usepackage{amssymb}
\usepackage{epsf}
\usepackage{epsfig}
\usepackage{graphicx}
\usepackage[english]{babel}
\usepackage[latin1]{inputenc}
\usepackage{subfigure}
\usepackage{color}
\usepackage{wrapfig}

\usepackage{a4wide}

\newcommand{\xianiso}{\xi_{\text{aniso}}}
\newcommand{\gammaaniso}{\gamma_{\text{aniso}}}
\newcommand{\Lhat}{\hat{\Lambda}}
\newcommand{\Phihat}{\hat{\Phi}}
\newcommand{\chihat}{\hat{\chi}}
\newcommand{\Rhat}{\hat{R}}
\newcommand{\kappahat}{\hat{\kappa}}
\newcommand{\xhat}{\hat{x}}
\newcommand{\yhat}{\hat{y}}

\newcommand{\what}{\hat{w}}
\newcommand{\MetG}{\textit{M-G}}
\newcommand{\Methom}{\textit{M-h}}
\newcommand{\Metmod}{\textit{M-m}}
\newcommand{\CA}{\textit{CA}}
\newcommand{\MetGspace}{\textit{M-G }}
\newcommand{\Methomspace}{\textit{M-h }}
\newcommand{\Metmodspace}{\textit{M-m }}
\newcommand{\CAspace}{\textit{CA }}
\newcommand{\nthermal}{n_{\text{thermal}}}
\newcommand{\nsweeps}{n_{\text{sweeps}}}
\newcommand{\nconfig}{n_{\text{config}}}

\newcommand{\Real}{\text{Re} \,}
\newcommand{\Imag}{\text{Im} \,}
\renewcommand{\vec}[1]{{\bf #1}}
\newcommand{\eqb}{\begin{equation}}
\newcommand{\eqe}{\end{equation}}
\newcommand{\dmb}{\begin{displaymath}}
\newcommand{\dme}{\end{displaymath}}

\newcommand{\eab}{\begin{eqnarray}}
\newcommand{\eae}{\end{eqnarray}}

\newcommand{\be}{\begin{equation}}
\newcommand{\ee}{\end{equation}}

\begin{document}
\begin{center}
\begin{Large}
{\bf Scalar Field Theory with a Non-Standard Potential } \\
\end{Large}

\medskip
Sebastian Scheffler\\
\textit{Institut f\"ur Kernphysik, Technische Universit\"at Darmstadt, Schlossgartenstr. 9, 64285 Darmstadt, Germany}\\
\medskip
Ralf Hofmann \\
\textit{Institut für Theoretische Physik, Universität Karlsruhe (TH), Kaiserstr. 12, 76128 Karlsruhe, Germany }\\
\medskip
Ion-Olimpiu Stamatescu\\
\textit{Institut f\"ur Theoretische Physik, Universit\"at Heidelberg, Philosophenweg 16, 69120 Heidelberg, Germany}\\
\end{center}

\begin{abstract}
We study the phase structure of a 4D complex scalar field theory with a potential $V(\Phi) = \vert \,\frac{\Lambda^3}{\Phi} - \Lambda \Phi\,
\vert^2$ at zero and at finite temperature. The model is analyzed by 
mean-field and Monte Carlo methods. 
At zero temperature the theory falls in the
universality class of the 4D Ising model when varying $\Lambda$. 
The situation is less clear-cut for variations w.r.t. $\Lambda$ at large
temperatures and variations w.r.t. temperature at a given value 
of $\Lambda$. We observe temperature 
independence of the mass of the first excitation. 
\end{abstract}

\section{Introduction}\label{Introduction}

\begin{figure}[t!]
\begin{center}
\epsfig{file = 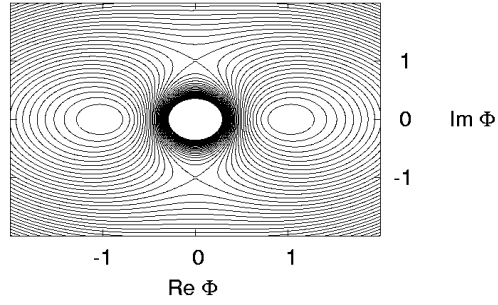  ,width = 10cm}
\caption{A contour plot of the potential Eq.~\eqref{Potential} for $\Lambda = 1.0$. The axes are scaled in units of $\Lambda$. Note the minima at $\Phi = \pm \Lambda$ on the real axis, the saddle points at $\Phi = \pm i \, \Lambda $ and the pole at $\Phi = 0$.}
\label{fig-plot-potential}
\end{center}
\end{figure}

The most striking success of the theory of critical phenomena is 
its ability to explain the 
coincidence of critical exponents in physical 
systems that, at first sight, seem to be vastly different
\footnote{Confer e. g. \cite{Fisher} for a review on this subject.}. The
behavior in the vicinity of continuous phase transitions 
can be explained in terms of just a few fundamental properties 
of the system under consideration: The 
dimension of spacetime and the symmetries of the associated 
Lagrangian. As a consequence, 
the tremendous variety of physical systems 
exhibiting critical behavior is classified by this universality.\\ 
It thus is of a certain interest to study the
nature of phase transitions in a system whose 
Lagrangian possesses unusual symmetries. Consider the 
field theory defined on four dimensional Euclidean spacetime by the Lagrangian density 
\begin{equation}\label{Lagrangian}
\mathcal{L}(\Phi, \partial_{\nu} \Phi) \; = \; \frac{1}{2}\lbrace (\partial_{\mu} \Phi^*)(\partial_{\mu} \Phi ) + V( \Phi )  \rbrace
\end{equation}
where the potential for the complex scalar field $\Phi$ is given as 
\begin{equation}\label{Potential}
\begin{split}
V(\Phi) \; & := \; \Bigl|  \frac{\Lambda^3}{\Phi} - \Lambda \Phi  \Bigr|^2  \\
& = \Lambda^2 \, | \Phi |^2 \: + \: \frac{\Lambda^6}{ | \Phi |^2 } - 2 \Lambda^4 \frac{ ( \text{Re}\, \Phi )^2 - ( \text{Im}\, \Phi )^2 }{  | \Phi |^2   }
\end{split}
\end{equation}
with $\Lambda$ being a constant of dimension mass. This model has been
proposed in the framework of a more general effective theory as described in~\cite{Ralf2, Hofmann2005C, Hofmann2007} and we refer the reader to these papers for details. A contour plot of the potential \eqref{Potential} is shown in Fig.~\ref{fig-plot-potential}. Notice that 
$V(\Phi)$ is symmetric under both of the following discrete
transformations: $\Phi \mapsto - \Phi$ and $\Phi\mapsto \Phi^*$ or,
equivalently, $\text{arg} (\Phi) \mapsto \text{arg} (\Phi) + \pi $ 
and $\text{arg}(\Phi) \mapsto - \text{arg} (\Phi)$. 
Notice also that the shape of the potential 
suggests the existence of phase transitions. 

In this work we aim at obtaining the following 
information about the theory defined by the Lagrangian \eqref{Lagrangian}: 
First, we are interested in resolving the phase structure of 
the model in the parameter $\Lambda$ and the temperature $T$. 
The latter will be introduced by an anisotropic coupling $\gamma$ in a 
lattice model. That is,
we would like to identify the line in the $\Lambda$-$\gamma$ plane which separates 
the phase with dynamically broken $\mathbb{Z}_2$- symmetry from the
phase where this symmetry is respected by the ground state. 
Second, we intend to investigate the order of the phase transition 
and - if a continuous phase transition occurs - to determine its critical 
exponents. Finally, we also determine the 
mass of the first excitation at both zero and finite temperature. 
This is used to relate the bare asymmetry parameter $\gamma$ to 
temperature. 

The paper is organized as follows. In section \ref{ModelDefinition} 
we define the lattice version of the continuum field theory given by the 
Lagrangian density \eqref{Lagrangian} and discuss the limiting cases
$\Lambda \rightarrow 0$ and 
$\Lambda \rightarrow \infty$. Subsequently, we carry out a 
mean-field analysis for 
the zero temperature case in section \ref{MFA}, which allows us to gain
first 
insights into the phase structure of the theory. Section \ref{MCstudy} 
is concerned with a Monte Carlo study of the model, and it constitutes 
the central part of this work. We first design and test suitable 
Monte Carlo algorithms in section
\ref{MC-algorithms}. In the remainder of section \ref{MCstudy} 
we obtain the 
phase diagram, determine the masses
and the physical 
anisotropy, and we carry out a finite-size scaling analysis to determine
the critical 
exponents. Finally, we summarize our results in section \ref{Summary}. \\

\section{Lattice version of the model}\label{ModelDefinition}

We discretize the field theory defined by
\eqref{Lagrangian} on a four dimensional lattice in
Euclidean spacetime. The field at site $n$  is referred to as $\Phi_n$ where $n$ is a lattice point. The lattice spacings along the
temporal and spatial axes are denoted by $a_{\tau}$ and $a_s$,
respectively. We label unit vectors by a hat, e. g. $\hat{\tau} \, , \, \hat{j}$, and their length is one in units of the respective lattice spacing. In particular, $n + \hat{j}$ is the nearest neighbor of site $n$ along the (spatial) $j-$ axis and similarly for $n + \hat{\tau}$. Following common conventions, Roman indices refer to spatial coordinates while Greek indices run over all coordinates of the compactified Euclidean spacetime including the temporal axis.\\
The discretized action then reads 
\begin{equation}\label{LatticeLagrangian}
\begin{split}
S[\Phi]&= \frac{a_s^4}{2 \gamma} \, \sum_n \, \Bigl\{\, -
\frac{\gamma^2}{a_s^2} \bigl( \Phi_{n+\hat{\tau}} \Phi_n^* + \Phi_n
\Phi^*_{n + \hat{\tau}} \bigr) - \frac{1}{a_s^2} \, \sum_j \, \bigl(
\Phi_{n + \hat{j}} \Phi_n^* + \Phi_n \Phi_{n + \hat{j}}^*  \bigr) \\
& \frac{2}{a_s^2} ( 3 + \gamma^2) \vert \Phi_n \vert^2  + V(\Phi_n) \, \Bigr\} \,.
\end{split}
\end{equation} 
Here we have introduced an asymmetry parameter 
$\gamma$ (also referred to as bare anisotropy) between temporal and spatial couplings. This allows us to study the model at
finite temperature along similar lines as in \cite{Karsch1,Karsch-Nucu}. We frequently decompose the
field variables 
into their real and imaginary parts according to $\Phi_n \equiv x_n + i
y_n$. It will turn out to be convenient to define the following quantities: 
\begin{equation}\label{Defs4LattAct}
\begin{split}
\Sigma_n & := \frac{\gamma}{a_s^2} \, ( \Phi_{n + \hat{\tau}} + \Phi_{n - \hat{\tau}} ) + \frac{1}{a_s^2} \, \sum_j \, \bigl( \Phi_{n + \hat{j}} + \Phi_{n - \hat{j}} \bigr) \\
\kappa & := \frac{1}{2} \, \bigl[  \Lambda^2 + \frac{2}{a_s^2} \, ( \gamma^2 + 3 ) \bigr] \\
w(x_n, y_n) & := - \frac{a_s^4}{\gamma} \, \Bigl\{ \frac{1}{2} \frac{\Lambda^6}{x_n^2 + y_n^2} - \Lambda^4 \, \frac{x_n^2 - y_n^2}{x_n^2 + y_n^2} \, \Bigr\} \; .
\end{split}
\end{equation}
At some points it will be convenient to write $\Phi_n$ as  $\Phi_n \equiv R_n e^{i \Theta_n}$. In terms of these variables the lattice action reads 
\begin{equation}\label{LattActionPolarCoords}
\begin{split}
S = \frac{a_s^2}{\gamma} & \, \sum_n \, \Bigl\{  - R_n \, \Bigl[ \gamma^2 R_{n + \hat{\tau}} ( \cos \Theta_{n + \hat{\tau}} \cos \Theta_n + \sin \Theta_{n + \hat{\tau}} \sin \Theta_n ) \\
& + \sum_j \, R_{n+\hat{j}} ( \cos \Theta_{n + \hat{j}} \cos \Theta_n  + \sin \Theta_{n + \hat{j}} \sin \Theta_n ) \Bigr] \\
& + ( 3 + \gamma^2 + \frac{a_s^2\Lambda^2}{2} ) R_n^2  + a_s^2 \frac{\Lambda^6}{2 R_n^2} - a_s^2 \Lambda^4 \cos( 2 \Theta_n) \Bigr\} \, .
\end{split}
\end{equation}
Also, we adopt the following notation: 
For each dimensionful physical quantity (as for instance $\Phi$,
$\Lambda$ etc.) we refer to the corresponding dimensionless
quantity as measured in appropriate powers of the spatial lattice
spacing $a_s$ by the same symbol endowed with a hat (e .g. $\Phihat$,
$\Lhat$ etc.).\\ 
  
Before addressing the dynamics of the model we analyze 
its behavior in the limits of weak and strong coupling $\Lambda$ at zero temperature ($\gamma = 1$). For $\Lambda \rightarrow 0$  we find 
\begin{equation}\label{Limit-lambda2zero}
S \; = \; \sum_n \Bigl( 4 + \frac{\Lhat^2}{2}  \Bigr) \hat{R}_n^2 \; - \; \sum_{< l, n > } \: \Rhat_{l} \Rhat_{n} \, \vec{S}_{l} \cdot \vec{S}_{n} \; + \; \mathcal{O}(\Lhat^4) \; ,
\end{equation}
where we have defined $\vec{S}_n := (\, \cos \, \Theta_n \, , \, \sin \, \Theta_n \,)$. The sum with the subscript $ < l, n >$ means summation over all pairs of nearest neighbors. This expression resembles the action of a free theory with a mass term.\\
 
In the strong coupling limit $\Lambda \rightarrow \infty$ all field configurations 
which do not have the field sitting in one of the two minima of the 
potential  at $\pm \Lambda$ on the real axis are strongly suppressed. The remaining degree of freedom is the sign of the real part 
of the field, that is $s_n := \text{sign} ( \,\Real \Phihat_n)$. Its 
dynamics is governed by lowest-order terms in $\Lambda$ in the Euclidean
action which read 
\begin{equation}\label{Limit-lambda2infinity}
S_{\Lambda \rightarrow \infty } := - \, \Lhat^2 \, \sum_{< l, n >} \, s_{l} \, s_{n}\,, 
\end{equation}    
Eq.~\eqref{Limit-lambda2infinity} is nothing but the 
familiar Ising model in four dimensions.

\section{Mean-field analysis}\label{MFA}

In this section we  study the model at hand in the mean-field
approximation (MFA). We restrict our analysis to the isotropic case
$\gamma = 1$. We find that the relevant symmetry is the 
$\mathbb{Z}_2$-symmetry relating the real minima of
the potential. A phase transition occurs
at a critical value $\Lhat_C$  of the lattice coupling constant 
above which
this symmetry is dynamically broken. In the following we shall stick to the ``old fashioned'' MFA (see, e.g.
\cite{stanley}) which is more intuitive. 

Let us consider the contribution to the overall action of the fixed field variable $\Phi_n$. From Eq.~\eqref{LattActionPolarCoords} one finds that for $\gamma = 1$ this contribution is given as 
\begin{equation}\label{Action-1site-polar-coords}
\begin{split}
S(\Phihat_n) \; & = \; - \Rhat_n \, \sum_{\hat{\mu}} \, \Bigl[ \cos \Theta_n \, \Bigl( \Rhat_{n +\hat{\mu}} \cos \Theta_{n+\hat{\mu}}  +  \Rhat_{n - \hat{\mu}} \cos \Theta_{n -\hat{\mu}}  \Bigr) \\
 & \hspace{0.5cm}   + \sin \Theta_n \, \Bigl( \Rhat_{n +\hat{\mu}} \sin \Theta_{n+\hat{\mu}} + \Rhat_{n - \hat{\mu}} \sin \Theta_{n -\hat{\mu}}  \Bigr)  \Bigr] \\
 & \hspace{0.5cm}  + \underset{ \hspace{1cm} =: \, \tilde{V}(\Rhat_n) }{\underbrace{\bigl( d + \frac{\Lhat^2}{2}  \bigr) \, \Rhat_n^2 + \frac{\Lhat^6}{2 \, \Rhat_n^2} }}- \Lhat^4 \cos( 2 \Theta_n) \; ,
\end{split}
\end{equation}
where $d$ denotes the dimension. If not stated otherwise we set $d=4$.

\begin{figure}[t!]
\begin{center}
\epsfig{file=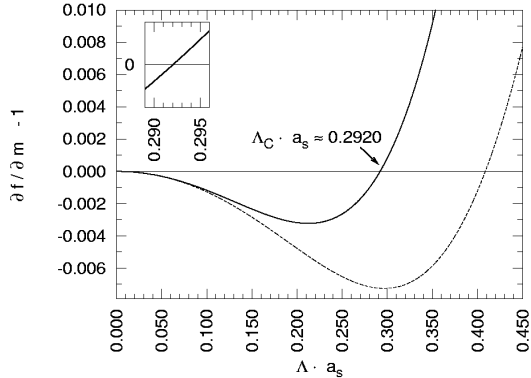, width =8cm }
\caption{This figure illustrates the determination of $\Lhat_C$ 
from the criterion $ \frac{ \partial f}{\partial m}(\Lhat, m) - 1 \overset{!}{=} 0 $, 
the inset shows the vicinity of the zero in more detail ($\omega=0$,
 solid curve). The dashed curve is the result for $\omega=\frac{\pi}{2}$, 
showing a later onset.}
\label{fig-MFA-deriv}
\end{center}
\end{figure}

The mean-field approximation amounts to decoupling each field variable
$\Phihat_n$ from its neighbors 
by replacing each of the $2 \cdot d$ adjacent fields $\Phihat_{n
  \pm \hat{\mu} }$ by a mean-field $m e^{i\omega}$, thus 
(dropping the indices '$n$') Eq.~\eqref{Action-1site-polar-coords} becomes:
\begin{equation}\label{MFA-2}
\begin{split}
 S(\Phihat) \; & = \; - 2 d m \Rhat \, ( \cos \Theta \cos \omega + \sin \Theta \sin \omega  )  - \Lhat^4 \cos ( 2 \Theta ) + \tilde{V}(\Rhat) \\
 & = \; - 2 d m \Rhat \cos(\Theta - \omega) -  \Lhat^4 \cos(2 \Theta) + \tilde{V}(\Rhat)  \; .
\end{split}
\end{equation}
 As a result the partition
function factorizes and we only have to deal with a single integral:
\begin{equation}\label{MFA-3}
\begin{split}
 Z(\Lhat, m ) \; & = \; \int d^2 \Phihat\, e^{-S(\Phihat)} \\
 & =  \; \int_0^{\infty} dR \int_0^{2 \pi} d \Theta \, R \exp \, \Bigl\{ + 2 d m R \cos(\Theta - \omega ) + \Lhat^4 \cos(2 \Theta) - \tilde{V}(R)  \Bigr\} \; .
\end{split}
\end{equation}
Selfconsistency then requires the expectation value of $\Phihat$ 
to equal the mean field, hence:
\begin{equation}\label{MFA-4}
\begin{split}
m\,e^{i\omega} = \langle \, \Phihat \, \rangle 
&=  \frac{1}{Z(\Lhat, m )} \, \int_0^{\infty} R\,dR \int_0^{2 \pi} d \Theta \, e^{-S(R, \Theta)} \, R \, e^{i \Theta } \\
& =: \, \varphi(\Lhat, m) \; .
\end{split}
\end{equation}
This mean-field equation always has the trivial solution  
$m = \langle \, \Phihat \, \rangle =0$. It can be shown from 
the symmetries of the integral
that non-trivial solutions require $\omega =0 $ (or $\pi$) and 
$\omega =\frac{\pi}{2} $ (or $-\frac{\pi}{2}$). For
small $ \Lhat$ only the trivial solution $m=0$ exists. With increasing 
$ \Lhat$ a non-trivial solution appears at some value $ \Lhat_C$, above 
which $\Phihat $ develops a non-zero expectation value $m(\Lhat) \ne 0$.
Since this solution has a smaller free energy than the trivial
 one\footnote{The trivial solution has in fact infinite free energy, but this 
is an artefact of the MFA for a potential singular at 0. } it 
replaces the latter, triggering a phase transition.  
Some detail is given in appendix A.1. It turns out that the 
solution with $\omega = \frac{\pi}{2}$ sets in at a larger $ \Lhat$
than the one with $\omega=0$ and has a larger free energy than the 
latter, therefore it does not take over. This is physically sensible,
since it corresponds to a mean field sitting on the saddle points
of the potential and not at the minima. 
In the following we shall therefore take $\omega=0$. See also Fig.~
\ref{fig-MFA-deriv}.

We define
\begin{equation}\label{Def-F_n}
F_n(\Lhat, m) := \frac{\partial^n Z}{\partial m^n}(\Lhat, m) \; 
\end{equation}
which can be obtained analytically (see appendix A.1). Then
\begin{equation}\label{MFA-f}
\varphi(\Lhat, m) \; = \; \frac{1}{2 d} \, \frac{F_1(\Lhat, m) }{F_0(\Lhat , m) }\; .
\end{equation}
The onset of the symmetry breaking solutions is given by:
\begin{equation}\label{MFA-7}
 \frac{\partial \varphi}{\partial m}(\Lhat_C, 0) = 1 \; .
\end{equation}
This is displayed in Fig.~\ref{fig-MFA-deriv}. For later 
comparison to the Monte Carlo results we quote the numerical value 
\begin{equation}\label{MFA-Lambda_C}
\Lhat_{C, \text{MFA}} \; \simeq \; 0.2920 \; .
\end{equation}

\begin{figure}[t!]
\begin{center}
\epsfig{file=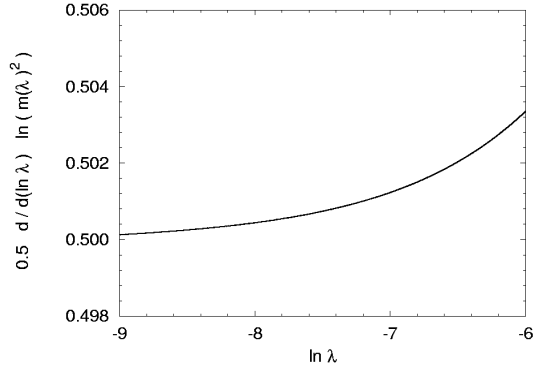, width=8cm}
\caption{Determination of the critical exponent $\beta$ from the MFA. The plot shows $ \frac{1}{2} \, \frac{d}{d (\ln \lambda)} \, \ln ( m(\lambda)^2 ) $ plotted versus $\ln \lambda$.}
\label{fig-MFA-beta}
\end{center}
\end{figure}

It is straightforward to obtain further quantities, for
example the critical exponent $\beta$ which governs the 
behaviour of the field expectation value near the threshold:
\begin{equation}\label{Def-beta}
\langle \,  \Phi \, \rangle \sim \Bigl( \frac{\Lambda - \Lambda_C}{\Lambda} \Bigr)^{\beta} \;  \text{for} \; \Lambda > \Lambda_C \;.
\end{equation}
For this we expand (\ref{MFA-f}) into a Taylor series in $m$. 
Making use of relation \eqref{Def-F_n} and noticing that
$F_n(\Lhat,0) = 0$ for odd $n$, 
we arrive at
\begin{equation}\label{expand-m-1}
m = \frac{1}{2d} \, \frac{m F_2(\Lhat, 0) + \frac{1}{6} m^3 F_4(\Lhat, 0) + \mathcal{O}(m^5)}{F_0(\Lhat,0) + \frac{m^2}{2} F_2(\Lhat, 0) + \mathcal{O}(m^4)} \; .
\end{equation}
Equating the terms up to fourth order in $m$ one readily finds
\begin{equation}\label{expand-m-2}
m^2 \:  = \: \frac{F_2(\Lambda, 0) - 2 d F_0(\Lambda, 0) }{d F_2(\Lambda, 0) - \frac{1}{6} \, F_4(\Lambda, 0) } \; .
\end{equation}
On Fig.~\ref{fig-MFA-beta} we plot the derivative of $\ln (
m(\lambda)^2 ) / 2$ as computed from Eq.~\eqref{expand-m-2} versus
 $\ln( \lambda)$ with:  
\begin{equation}
\lambda := ( \Lambda - \Lambda_C) / \Lambda_C
\;.
\end{equation}
In the limit $\lambda \rightarrow 0$ we obtain $\beta=0.5$ which is
what should be expected for the universality class of the four 
dimensional Ising-model.\\

\section{Monte Carlo study}\label{MCstudy}

In this section we study the model by means of Monte Carlo
simulations. We first outline algorithms to sample 
field configurations and discuss their performance. 
In section \ref{MC-phase-diag} we present a phase diagram of the model.
 Subsequently, we analyze the mass gap of the theory at zero and finite 
temperature and estimate the relation between the bare coupling
anisotropy 
$\gamma$ and the renormalized, physical anisotropy $\xianiso$ in 
section \ref{Mass-gap}. Finally, section \ref{FSA} presents a
finite-size 
scaling analysis where we determine $\Lambda_C$, several 
critical exponents, and the order of the phase transition. \\

\subsection{Monte Carlo algorithms for the model}\label{MC-algorithms}

Starting from Eqns.~\eqref{LatticeLagrangian} and \eqref{Defs4LattAct}, 
the contribution to the overall action of a fixed site $n$ can be written as 
\begin{equation}\label{Action-1site}
S(\Phi_n) = \frac{\kappa}{\gamma} \,\bigl( x_n - \frac{ \Real \Sigma_n}{2 \kappa} \bigr)^2 + \frac{\kappa}{\gamma} \, \bigl( y_n - \frac{ \Imag \Sigma_n}{2 \kappa} \bigr)^2 - w(x_n, y_n) \, .
\end{equation}
That is, the action can be decomposed as 
\begin{equation}\label{decompose-S}
S[\Phi] = S(\Phi_n) + \, \bigl( \, \text{terms independent of} \, \Phi_n \, \bigr) \, .
\end{equation}
First, we implement a standard local Metropolis algorithm \cite{Metro}. From Eq.~\eqref{Action-1site} one sees that the probability distribution for a single field variable $\Phihat_n$ in an otherwise fixed field configuration is given by\footnote{ $w(x,y)$ has been defined in Eq.~\eqref{Defs4LattAct}.}
\begin{equation}\label{PD1site}
\begin{split}
p(\Phihat_n) & \propto e^{-S(\Phihat_n)} \\
 & \propto \exp \bigl\{ - \frac{\kappahat}{\gamma} \, (\xhat_n - \xhat_0 )^2 \bigr\} \, \exp \bigl\{ - \frac{\kappahat}{\gamma} \, (\yhat_n - \yhat_0 )^2 \bigr\} \, \exp \bigl\{  \, \what(\xhat_n, \yhat_n)    \bigr\} \, .
\end{split}
\end{equation} 
A local Metropolis algorithm can be implemented by generating proposals $\xhat^\prime$ and $\yhat^{\prime}$ for the new real 
and imaginary part of $\Phihat_n$ from Gaussian
distributions with the parameters 
found in eq.\,(\ref{PD1site}). The trial 
field variable $\Phihat^\prime_n \equiv \xhat^\prime + i \yhat^\prime$
is accepted with probability $\min \, \bigl\{ \, 1\, ,\,  \exp [ \,
\what(\xhat^\prime, \yhat^\prime) - \what(\xhat_n, \yhat_n) \, ] \,
\bigr\} \,$. We shall refer to this algorithm as \MetGspace in the
following.\\
A different way to set up a local
Metropolis sampler is to generate trial variables $\xhat^\prime$ and
$\yhat^\prime$ from a homogeneous distribution on the intervals $[\xhat_0 -a_x, \xhat_0 + a_x]
$ and $[\yhat_0 - a_y, \yhat_0 + a_y]$, respectively. Here $a_x$ and
$a_y$ are constants which can be chosen arbitrarily. 
Given the proposal  $\Phihat^\prime_n = \xhat^\prime + i \yhat^\prime$
one then accepts $\Phihat^\prime$ with the probability $ \min \lbrace \,
1 \, , \, \exp \, [ \, S_E( \, \Phihat_n \, ) - S_E( \, \Phihat^\prime
\, )\, ] \, \rbrace   $. 
We dub this procedure \Methom. The efficiency of this algorithm will
strongly depend on the choice of the interval 
widths $a_x$ and $a_y$.

It is expected that the algorithm \Methomspace can be rendered more
efficient by implementing 
it as 'modified Metropolis algorithm' \cite{Creutz2} which performs several subsequent updates of the same site during each sweep over the lattice. We will test this procedure as well and refer to it as \Metmod. 

Since we anticipate the occurrence of a continuous phase transition (cf. section \ref{MFA}) we also implement a Cluster algorithm in order to reduce the possible impact of critical slowing down. We adapt the methods discussed in \cite{Brower, Sok1} to our model. Our implementation of the cluster algorithm is outlined in detail in the appendix \ref{CA-App}. Given that the cluster algorithm only refreshes a subset of the degrees of freedom it has to be combined with one of the local Metropolis samplers described in the beginning of this section. We have chosen the algorithm \MetGspace for this purpose and we refer to this combination of the cluster algorithm and \MetGspace as \CAspace in the following. \\

\begin{figure}[t!]
\begin{center}
\epsfig{file= 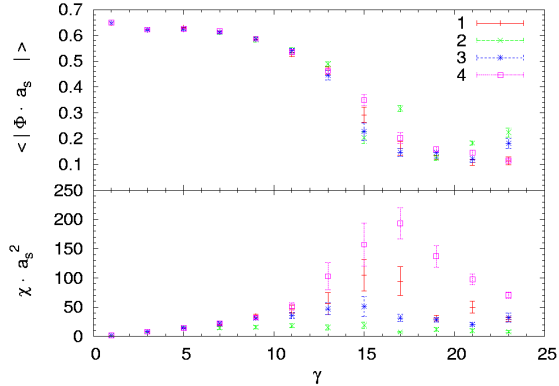, width = 8cm}
\caption{(Color online) Plots of the order parameter and the susceptibility as obtained using the algorithms discussed in the text. The numbers in the legend are related to the algorithms in the following way: (1) \MetG;  (2) \Methom; (3) \Metmod; (4) \CA. }
\label{fig-compare-alg}
\end{center}
\end{figure}
We now report on our practical experiences with the above algorithms. In order to compare the algorithms we have carried out simulations using each of the algorithms on a $12^4-$ lattice at fixed $\Lhat = 0.5$ varying the coupling anisotropy $\gamma$. After initializing the lattice in a disordered way ('hot start') 20000 thermalization sweeps were performed and 1000 configurations were subsequently analyzed, separated by 50 sweeps each. We have computed the expectation value of the field's modulus\footnote{Cf. \cite{Talapov-Bloethe} for a discussion of the validity of this procedure and its systematic errors.} $\langle \, | \Phihat | \, \rangle  := | \,\sum_n \, \Phihat_n \, | / V_4 $ and the corresponding susceptibility $\chihat = V_4 \bigl[ \langle \, | \Phihat \, |^2 \, \rangle - \langle \, | \Phihat \, | \, \rangle^2 \, \bigr]$ ($V_4$ denotes the four dimensional volume of the lattice.). The results are displayed in Fig.~\ref{fig-compare-alg}. As can be seen, there is a phase transition at $\gamma \simeq 14$ (This will be discussed in detail in section \ref{MC-phase-diag}.) Below this transition ($\gamma \lesssim 12$) the results obtained 
with all algorithms agree within the margins of error\footnote{These were estimated by means of the jackknife method.}. However, deviations exist for larger 
values of $\gamma$. 
We think that these are due to inefficiencies 
of some of the algorithms in the vicinity of the phase transition. The \Methom update performs particularly poorly in this range. The acceptance rate was very low and this could not be significantly improved by tuning the widths $a_x$ and $a_y$ of the proposal distribution. Its modified version \Metmodspace performs significantly better; indeed the data is in better agreement with the data generated by means of the other algorithms. We have decided to use  the algorithms \MetGspace and \CAspace exclusively in the remainder of the work because they perform better and do not require tuning the widths of the proposal distributions. \\
\begin{table}
\begin{center}
\begin{scriptsize}
\begin{tabular}[c]{|c|c|c|c|c|}
\hline
data displayed/quoted in & $\nthermal$ & $\nsweeps$ & $\nconfig$ & MC- alg. \\ \hline
Fig. \ref{fig-compare-alg} & 20000 & 50 & 1000 & various ones \\
Sec. \ref{MC-phase-diag}  & 10000 to 20000 & 50 & 1000 & \MetG \\
Tab. \ref{mass-results} & 200000 to 30000 & 50 & 2800 to 200000 (*) & \MetG \\
Tab. \ref{xi-results} & 20000 to 30000 & 50 & 8000 to 60000 (*) & \MetG  \\
Tab. \ref{finite-T-mass-results} & 10000 & 50 & 10000 ($\gamma=2$),  20000 ($\gamma=3$) & \MetG \\
Sec. \ref{FSA} & 10000 & 50 & 1000 & \CA \\
\hline
\end{tabular}
\caption{Summary of the statistics of the simulations in this work. (*): The number of configurations depends on the lattice size. - Confer the text for further explanations. }
\label{simulation-parameters}
\end{scriptsize}
\end{center}
\end{table}
We close this section by briefly describing some other technical details. In Tab.~\ref{simulation-parameters} an overview is given on the statistics and algorithms used to obtain the results quoted in this work. Here, $\nthermal$ denotes the number of thermalization sweeps that were carried out before observables were measured the first time after initializing the lattice. $\nsweeps$ stands for the number of updates of the entire lattice carried out between every two successive measurements. $\nconfig$ gives the number of configurations from which observables were calculated. We have tested ordered, disordered, and mixed phase initial configurations and checked that they yield the same results. We have always applied periodic boundary conditions.\\
The margins of error quoted in this work and displayed in the figures were obtained by means of the jackknife method if not stated otherwise. 

\subsection{Phase diagram of the theory}\label{MC-phase-diag}
\begin{figure}[t!]
\begin{center}
\epsfig{file = 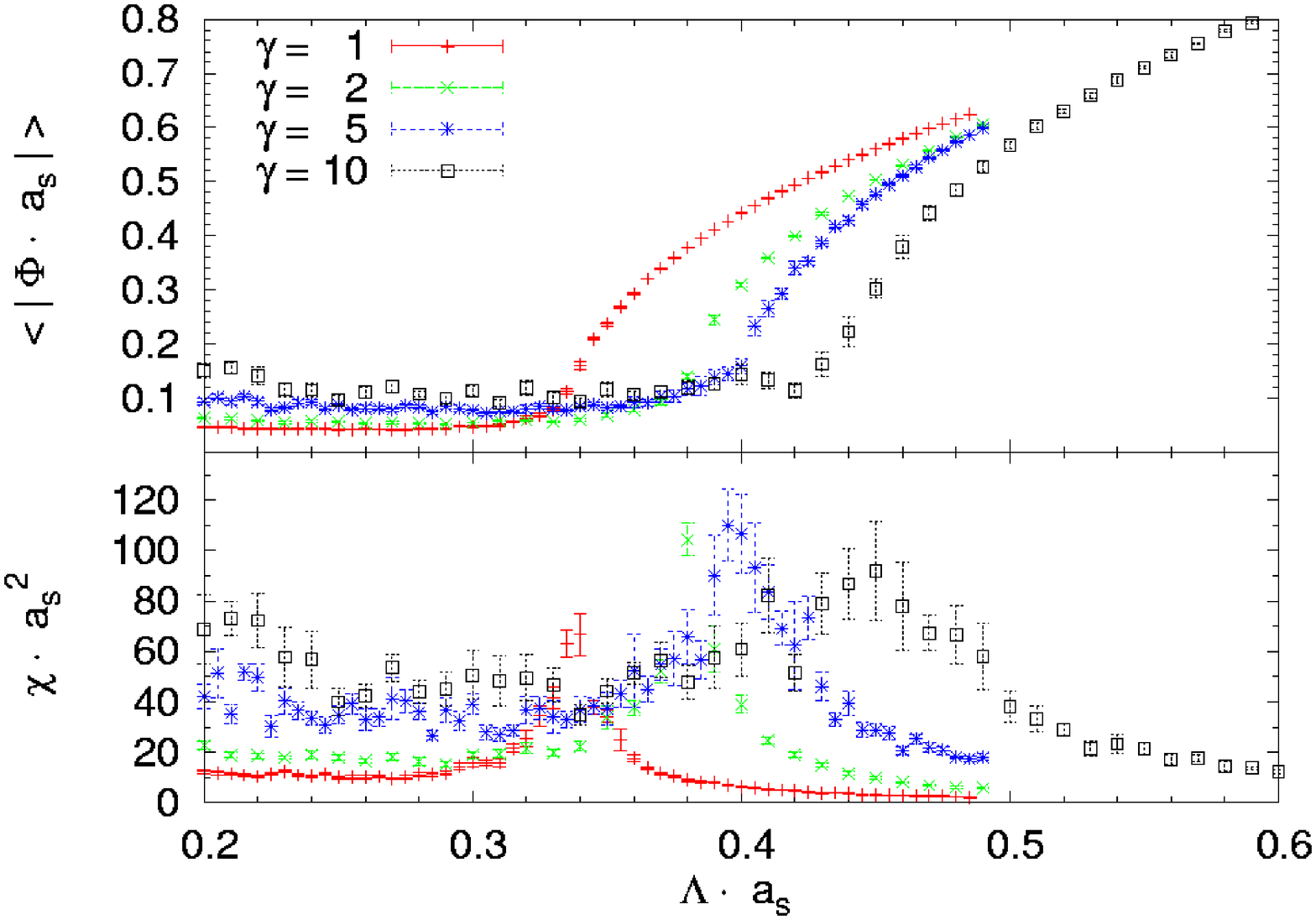, width = 8cm }
\caption{(Color online) Resolving the phase structure by running simulations cutting through the $(\gamma, \Lhat)$- plane. This figure presents results obtained at fixed values of $\gamma$. }
\label{fig-resolve-phase-structure}
\end{center}
\end{figure}
In this section we present a phase diagram of the model in the plane
spanned by the parameters $\Lhat$ and $\gamma$. From the mean-field analysis (cf. section \ref{MFA}) we expect
spontaneous symmetry breaking to occur within a certain range of this
plane. 

\begin{figure}
\begin{center}
\epsfig{file = 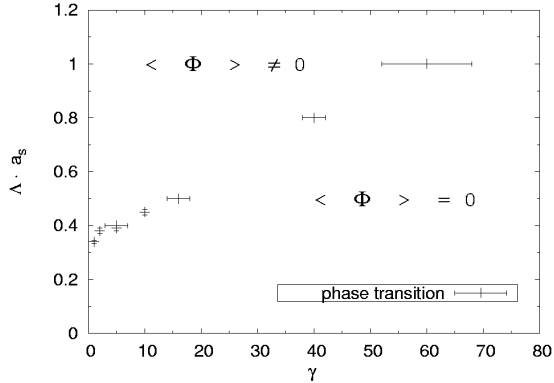, width = 8cm  }
\caption{Phase diagram of the model.}
\label{fig-phase-diag}
\end{center}
\end{figure}

The phase structure is resolved by running simulations varying one of the
parameters 
while keeping the other one fixed. 
Fig.~\ref{fig-resolve-phase-structure} shows results for the expectation value of the field (i. e. the order parameter for the system under consideration) and the susceptibility as obtained from simulations where the coupling $\Lhat$ was increased at different values  of the anisotropy parameter $\gamma$. The sudden increase in each of the curves displayed in the upper panel together with the simultaneous peak in the susceptibility provides strong indication for a phase transition. Given that the results for the order parameter $\Phihat$ appear to be continuous we suppose that the phase transitions are continuous\footnote{We will provide evidence for the continuous character of the phase transition at $\gamma = 1$ and $\gamma = 10$ in section \ref{FSA} where we carry out a finite-size scaling analysis.}. From the peaks of the susceptibilities we have read off the critical
values of the parameters and estimated their errors from the width of
the peaks. Plotting the phase transition points in the $(\gamma ,
\Lhat) \, $- plane yields Fig.~\ref{fig-phase-diag} which provides the
phase diagram of the model. Briefly summarizing its content in words,
one can say that there is a line of presumably second-order phase
transitions separating the symmetric phase at small $\Lhat$ and/or high
$\gamma$ from a symmetry broken phase which exists at larger values of
$\Lhat$ and not too high values of $\gamma$. Two features deserve to be
commented on. First, it appears as if there is a critical $\Lhat_C$
below which spontaneous symmetry breaking never occurs. This can be
inferred from the results of the simulations keeping $\gamma = 1.0 $
fixed\footnote{Physically, we interprete the case $\gamma = 1$ as the
  zero temperature limit while higher numerical values correspond to
  non-zero temperature. Thus we consider $1.0$ as
  the lowest attainable value of $\gamma$.}. Second, the numerical value
of this critical coupling can be roughly estimated as $\Lhat_C \approx
0.3$ which agrees remarkably well with the prediction from the mean
field analysis quoted in Eq.~\eqref{MFA-Lambda_C}\footnote{Also this
  issue is addressed in more detail in section \ref{FSA}.}. We notice that there are larger errors at high $\gamma$. This will be discussed in section \ref{Summary}.\\

We briefly comment on the role of the parameter $\Lambda = \Lhat \cdot
a_s^{-1}$ which has the dimension of a mass. When investigating the
phase structure of the lattice model we consider the dimensionless
$\Lhat$ as the relevant parameter. 
On the other hand, we can understand $\Lambda$ as a cutoff that 
is externally given. By virtue of the relation between $\Lambda$ and
$\Lhat$ we can then interpret $a_s$ 
as the spatial resolution or, equivalently, $a_s^{-1}$ as a momentum cutoff. \\

\subsection{Mass gap and physical anisotropy}\label{Mass-gap}
In this section we determine the mass gap of the theory at both zero and finite temperature and analyze the dependence of the physical anisotropy $\xianiso$ on the corresponding input parameter $\gamma$. \\
\begin{table}[t!]
\begin{center}
\begin{tabular}[c]{|c|c|c|c|c|}
\hline
$\Lhat$ &   Lattice   &  $m [a_s^{-1}]$ &   Eq.~\eqref{estimate-mass} \\ \hline
 0.5  & $ 8^4$  &      0.576  (2)  &     1.0 \\ \hline
 0.5  & $12^4$  &       0.558  (7)  &    1.0 \\ \hline
 0.5   & $16^4$  &       0.554  (6)  &   1.0 \\ \hline
 0.8  & $ 8^4$  &       1.139 (11)   &    1.6   \\ \hline
 1.0  & $ 8^4$  &       1.55   (3)   &    2.0   \\  \hline
 1.0  & $12^4$  &       1.57   (6)   &    2.0   \\   \hline
\end{tabular}
\caption{Results for the masses at zero temperature, i. e. $\gamma = 1$. The last column gives the tree level estimates in units of $a_s^{-1}$ as computed from Eq.~\eqref{estimate-mass}.}
\label{mass-results}
\end{center}
\end{table}
First, we study the mass gap of the theory at zero temperature, i. e. on
isotropic lattices where $N_{\tau} = N_x$ and $\gamma = 1$. Before
presenting our Monte Carlo results we provide an estimate for the mass
from the second derivative of the potential at its minima. Keeping in
mind that there is a factor $ 1 / 2 $ in front of $V(\Phi)$ in
Eq.~\eqref{Lagrangian} 
the estimate for the mass at tree level is 
\begin{equation}\label{estimate-mass}
m_0 = \sqrt{\frac{1}{2} \, \frac{\partial^2 V}{\partial (\Real \Phi)^2 }} = \sqrt{\frac{1}{2} \, \frac{\partial^2 V}{\partial (\Imag \Phi)^2 }} = 2 \Lambda \;.
\end{equation}
This will serve for later comparison to the numerical results. Notice
that $m_0$ is associated with the mass of topologically trivial
fluctuations of the field $\Phi$ about the minima
$\Phi_{\tiny\mbox{min}}=\pm\Lambda$ of its potential. The fact that $m_0$ is larger than
$|\Phi_{\tiny\mbox{min}}|$ implies that these fluctuations do not take
place. \\ 
We have numerically determined the lowest mass of an intermediate state 
by measuring effective masses computed from timeslice correlators. It
proved beneficial to run the calculations on lattices with different linear extensions because this allows to find a  reasonable trade-off between systematic errors (due to too short a time axis) and statistical errors (due to limited computer time when simulating larger volumes). Tab.~\ref{mass-results} presents the results for different choices of the coupling $\Lhat$. \\

In the next passage we present estimates for the physical anisotropy $\xianiso \equiv a_s / a_{\tau}$ resulting from a given bare anisotropic coupling $\gamma$. To this end, we proceed as follows. We require that the physical correlation length $\xi$ be the same when measured along temporal and spatial axes. That is,
\begin{equation}\label{xi-requirement}
\xi = a_s \xi_s \overset{!}{=} a_{\tau} \xi_{\tau} \,
\end{equation} 
which implies 
\begin{equation}\label{xi-aniso-from-a}
\xianiso = \frac{a_s}{a_{\tau}} = \frac{\xi_{\tau}}{\xi_s} = \frac{m_{\tau}}{m_s}
\end{equation}
where we have used $\xi_s = 1 / m_s$ (and respectively for $\xi_{\tau}$ ) in the last equality. $m_s$ and $m_{\tau}$ denote the masses measured along the respective axis in units of the inverse lattice spacing. \\
\begin{table}
\begin{center}
\begin{scriptsize}
\begin{tabular}[c]{|c|c|c|c|c|c|}
\hline
$\Lhat$ & $N_x$ & $N_{\tau}$ & $\gamma $ &  $d \:[a_s] $ & $\xianiso$  \\ \hline \hline
0.5   &   12   &   24     &  2.0  &     2    &     2.01 (4) \\ \hline
      &        &          &       &      3    &     1.92 (4) \\ \hline
      &        &          &       &          4    &     2.10 (11) \\ \hline
      &        &          &       &         5    &     2.2  (2)  \\ \hline
      &        &          &       &           6    &     2.1  (3)  \\ \hline \hline
0.5   &  8     &    24    &  3.0  &    2    &     2.98 (2)  \\ \hline
      &        &          &       &    3    &     2.98 (4)  \\ \hline
      &        &          &       &    4    &     3.00 (4)  \\ \hline \hline
0.8   &   12   &   24     &  2.0  &     2    &     1.97 (2)  \\ \hline
      &        &          &       &       3    &     1.90 (13) \\ \hline
      &        &          &       &      4    &     1.8 (4)   \\ \hline  \hline
0.8   &  8     &   24     &  3.0  &     2    &     2.88 (4) \\ \hline
      &        &          &       &      3    &     2.90 (12) \\ \hline
      &        &          &       &       4    &     3.2 (3)   \\ \hline \hline
1.0   &  8     &   16     &  2.0  &    2    &     1.88 (8)  \\ \hline
      &        &          &       &     3    &     2.0  (2)  \\ \hline
      &        &          &       &       4    &     1.4  (2)   \\ \hline \hline
1.0   &  8     &   24     &  3.0  &     2    &     2.70 (6) \\ \hline
      &        &          &       &       3    &     2.5  (2)  \\ \hline
      &        &          &       &        4    &     5 (4) \\ \hline
\end{tabular}
\caption{Results for the physical anisotropy. $d$ in the fifth column denotes the separation at which the effective mass has been evaluated.}
\label{xi-results}
\end{scriptsize}
\end{center}
\end{table}
We now make the assumption that $ \xianiso / \gamma = \mathcal{O}( 1
)$. This was found to hold in the case of $\phi^4$- theory \cite{MSchiestl}. Under this assumption the lattice is approximately isotropic when the edge lengths are chosen as $N_{\tau} = \gamma N_x$. This choice of the lattice extensions ensures that the impact of undesired finite temperature effects can be neglected. Using such a setup, we have measured effective masses as a function of distance and we have estimated $\xianiso$ from
\begin{equation}\label{estimate-xianiso}
\xianiso \simeq \frac{m_s (d)}{m_{\tau}(\gamma d)} \,.
\end{equation}
The purpose of evaluating $m_{\tau}$ at separation $\gamma d$ is to ensure that the contributions of higher excitations have decayed away to the same extent when comparing the two effective masses. Tab.~\ref{xi-results} summarizes our results from this procedure. As can be seen there, the physical anisotropy $\xianiso$ agrees with $\gamma$ within the margin of error. In particular, this implies for the temperature\footnote{We work in units where $k_B = 1$.} $T = \xianiso / ( a_s \, N_{\tau} ) = \gamma / ( a_s \, N_{\tau} )$. This shows that $\gamma$ is related to the physical temperature in a straightforward way and therefore it a posteriori justifies our use of $\gamma$ as a parameter in the phase diagram in section \ref{MC-phase-diag}. It has to be noted though that we have only checked the  relation between $\xianiso$ and $\gamma$ for $\gamma \leq 3$. We assume that we can extrapolate these results to larger values of $\gamma$. \\

Finally, we have studied the temperature dependence of the mass. To this end, we have applied the same procedure as in the zero temperature case but this time running the simulation on an anisotropic lattice with $\gamma > 1$. The results are displayed in Tab.~\ref{finite-T-mass-results}. 
By comparing this result to the corresponding one in Tab.~\ref{mass-results} 
one sees that the mass has slightly dropped but in general it does 
not exhibit a strong temperature dependence. We will comment on this in the conclusions (cf. section \ref{Summary}).

\begin{table}[t!]
\begin{center}
\begin{tabular}[c]{|c|c|c|c|c|c|}
\hline
$\Lhat$ &   $\gamma$  & $N_x$ & $N_{\tau}$ & $T \, [ a_s^{-1}]$ &  $m [a_s^{-1}]$   \\ \hline
0.5       &    2.0      &  12   &    12      & 1 / 6              &   0.53 (2)       \\  \hline
0.5       &    3.0      &  12   &    12      & 1 / 4              &   0.53 (2)       \\  \hline
\end{tabular}
\caption{Results for the masses at finite temperature.}
\label{finite-T-mass-results}
\end{center}
\end{table}

\subsection{Finite-size scaling analysis for the susceptibility}\label{FSA}
\begin{figure}[t!]
\begin{center}
\subfigure[$\gammaaniso = 1.0$ ]{\epsfig{file=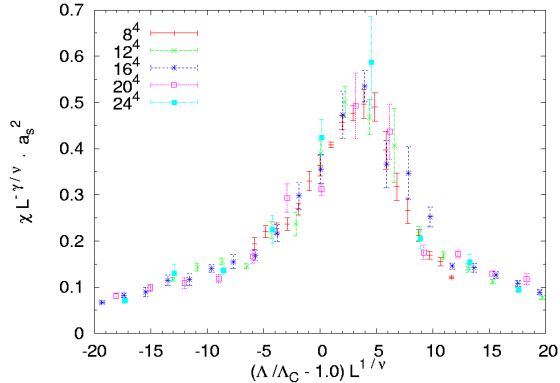, width = 8cm}}
\subfigure[$\gammaaniso = 10.0$ ]{\epsfig{file=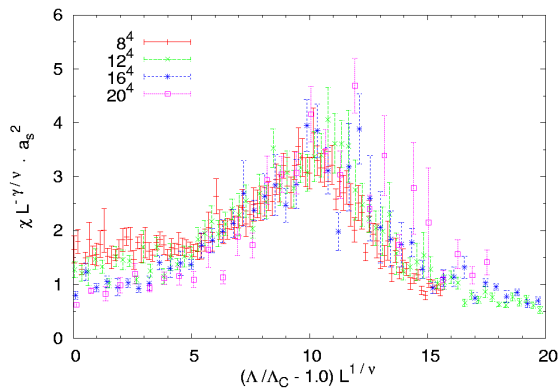, width = 8cm }}
\caption{(Color online) Test of the finite-size scaling. The values inserted for $\Lhat_C$, $\gamma$, and $\nu$ are the ones quoted in Tab.~\ref{FSA-results}.}
\label{fig-check-ffs}
\end{center}
\end{figure}

In this section we carry out a finite-size scaling analysis of the Monte Carlo data for the susceptibility $\chihat$. This will enable us to determine several critical exponents and hence the 
universality class of the model. 
Furthermore, we confirm that the phase transition when varying $\Lhat$ at fixed temperature is indeed continuous. \\
The data underlying the discussion in this section have been obtained from simulations on lattices of various linear extensions $L$. We have varied the coupling $\Lhat$ at several choices\footnote{Deviating from our conventions in the rest of this work we will denote the bare anisotropy parameter by $\gammaaniso$ in section \ref{FSA} in order to avoid confusion with the critical exponent $\gamma$.} of $\gammaaniso$. In analogy to the definition of a reduced temperature we define a reduced coupling $\lambda$ as 
\begin{equation}\label{Def-reduced-lambda}
\lambda  \, := \, \frac{\Lambda - \Lambda_C }{\Lambda_C }
\end{equation}     
where $\Lambda_C$ denotes the (a priori unknown) critical coupling\footnote{In fact, we should write this as $\Lambda_C(\gammaaniso)$ since we study the model at various, fixed choices of $\gammaaniso$. However, we will suppress the $\gammaaniso$- dependence in our notation for the sake of brevity. }. We denote the correlation length in the infinite volume limit by $\xi$. Then, the standard finite-size scaling approach states that the susceptibility $\chi_L(\lambda)$ as measured on a finite box of linear extension $L$ is given by 
\begin{equation}\label{FSA-Ansatz}
\chi_L(\lambda) = \xi^{\gamma / \nu } \chi_0( L / \xi ) \, .
\end{equation}
Here, $\gamma$ and $\nu$ denote the critical exponents such that in the thermodynamic limit $ \chi \sim \lambda^{-\gamma} $ and $\xi \sim \lambda^{ - \nu } $ in the vicinity of the critical point at $\lambda = 0$. $\chi_0$ is a continuous function which only depends on the ratio $ L / \xi $. Substituting the asymptotic behavior of $\xi$ into the last equation and making the ansatz
\begin{equation}\label{FSA-chi0-f}
\chi_0( L \lambda^\nu )\equiv \Bigl( \frac{L}{\xi} \Bigr)^{ \gamma / \nu } \, f( L^{ 1 / \nu} \, \lambda )
\end{equation}
with some continuous function $f$ one arrives at
\begin{equation}\label{FSA-Ansatz-2}
\chi_L(\lambda ) = L^{ \gamma / \nu } \, f( L^{1 / \nu } \, \lambda ) \, .
\end{equation}
Assuming an expression for $f$ one can fit data from lattices of different lengths using the last equation and thus one can determine the critical exponents and $\Lhat_C$ (The dependence of Eq.~\eqref{FSA-Ansatz-2} on $\Lhat_C$ enters via the definition of $\lambda$ in Eq.~\eqref{Def-reduced-lambda}.). \\
Introducing a set of six fit parameters $a_0$ to $a_5$ our ansatz for fitting $f$ reads
\begin{equation}\label{FSA-fit-ansatz-3}
y(\Lhat, L) \, := \, L^{ a_1 / a_2 } \frac{ a_3 }{\sqrt{1 + a_4^2[L^{1 / a_2}( \frac{ \Lhat}{a_0} -1 ) - a_5 ]^2}} \; .
\end{equation}
The significance of the fit parameters is as follows. $a_0$ corresponds to the value of the critical coupling $\Lhat_C$, $a_1$ and $a_2$ correspond to the critical exponents $\gamma$ and $\nu$, respectively. $a_5$ serves to take into account that the peak of the susceptibility as measured on a finite volume occurs slightly shifted away from its position at $\lambda =0$ in the thermodynamic limit. The parameters $a_3$ and $a_4$ are not of any immediate physical significance but serve to parameterize the function which resembles a Lorentz curve. \\
\begin{table}[t!]
\begin{center}
\begin{tabular}[c]{|c|c|c|c|c|c|}
\hline
$\gammaaniso$  & number of data points & $\Lhat_C$ & $\gamma$   & $\nu$      & $\chi^2$ /d.o.f. \\ \hline
1.0               &      51        & 0.3299 (3)  & 0.98 (3)   & 0.50   (2) & 2.2              \\ \hline
2.0               &     140        & 0.3719 (3)  & 0.924 (13) & 0.449 (11) & 1.7              \\ \hline
5.0               &     124        & 0.3810 (7)  & 0.84 (3)   & 0.421 (14) & 1.6              \\ \hline
10.0              &     103        & 0.352  (7)  & 1.05 (6)   & 0.67   (5) & 1.2              \\ \hline
\end{tabular}
\caption{Results for $\Lhat_C$ and the critical exponents $\gamma$ and $\nu$ at several values of $\gammaaniso$ as obtained from the finite-size scaling method. The last column gives the value of $\chi^2$ per degree of freedom. The margins of error are the estimates as obtained from the standard $\chi^2$- fit. Confer the text for a comment on systematic errors. }
\label{FSA-results}
\end{center}
\end{table}
Tab.~\ref{FSA-results} summarizes the results from fitting our Monte Carlo data while Fig.~\ref{fig-check-ffs} visualizes
the outcome of the finite-size scaling procedure for $\gamma = 1 $ and $ \gamma = 10$. Several things need to be commented on. First, the margins of error quoted in Tab.~\ref{FSA-results} are the statistical errors obtained from the $\chi^2$- fit\footnote{$\chi^2$ is the common merit function of the $\chi^2$- fit and not the square of the susceptibility $\chi$. Since there should be no risk of confusion we do not introduce an extra piece of notation to distinguish the two quantities.} in the common way; they appear to be underestimated, in particular for $\Lhat_C$. We think that there are systematic errors that have not been taken into account in Tab.~\ref{FSA-results} which affect the fit. 
This is because the statistical errors of the Monte Carlo data are largest in the vicinity of the critical point where the finite-size scaling method should make the  curves collapse. Therefore, the $\chi^2$- fit gives a lower weight to the data that are more relevant for physical reasons. It is difficult to 
estimate the numerical size of this effect.\\
Comparing the quality of the fit at the different values of $\gammaaniso$ (cf. Fig.~\ref{fig-check-ffs}), one sees that it works better at smaller $\gammaaniso$. In the limit of high temperature (Fig.~\ref{fig-check-ffs}(b) ) the statistical errors are significantly larger. In particular, we think that one should be a bit careful about the estimate for $\Lhat_C(\gammaaniso = 10) \approx 0.35 $ which appears unreliable when compared to the value read off from Fig.~\ref{fig-resolve-phase-structure}. \\
Giving an interpretation of these results, we first note that the critical exponents $\gamma$ and $\nu$ obtained at $\gammaaniso \lesssim 2 $ agree well with the ones for the Ising model in four dimensions. Thus we deduce that the model lies in the same 
universality class as the 4D-Ising model for these choices of the parameters. At this point it is necessary to discuss the non-renormalizability of the potential \eqref{Potential}. From an expansion about one of its minima\footnote{Due to the pole at $\Phi=0$ such an  expansion will only be valid in a sufficiently small vicinity of the respective minimum.} one finds that the potential contains contributions up to arbitrary order in the real and imaginary part of $\Phi$. The corresponding coupling constant for a contribution of order $n$ is proportional to $\Lambda^{-(n-4)}$. The terms with a coupling of negative mass dimension are irrelevant for the critical properties of the model \cite{Fisher}.  The only relevant and marginal contributions are the terms of order $\Phi^2$, $\Phi^3$ and $\Phi^4$ and the critical exponents should be  completely determined by them. \\
 The data at higher temperatures is not interpreted in as
 straightforward a way. From Tab.~\ref{FSA-results} one sees that the
 critical exponent $\nu$ shows a noticeable increase when $\gammaaniso$
 is raised to  $\gammaaniso = 10$ after dropping at intermediate
 values. In the limit of high temperature (i. e. $\gammaaniso = 10$),
 the numerical value of $\nu$ is compatible with the corresponding
 critical exponent of the three dimensional Ising model whose value is
 $0.630 \pm 0.0015$ \cite{Zinn-Justin}. This observation might hint at
 the dimensional reduction of the four dimensional field theory to a
 three dimensional one in the limit of infinite temperature (or
 equivalently $\gammaaniso \rightarrow \infty$). Indeed, it was 
found in \cite{Nucu-vL, O-Connor-1, O-Connor-2, Liao-Strickland} that a
dimensional crossover 
can be observed in the critical properties of a field theory when the
temperature 
is increased from $T=0$ towards $T \rightarrow \infty$. \\

From the values quoted in Tab.~\ref{FSA-results} we also infer that the phase transition is always a 
continuous one. For a first-order phase transition the susceptibility at the critical point 
diverges proportional to the volume \cite{Fisher-Berker,Cardy,Binder-Landau}. From Eq.~\eqref{FSA-Ansatz-2} it follows that $\chi \sim (L / \xi)^{ \gamma / \nu} $. Given that our data  is certainly incompatible with $ \gamma / \nu = 3 $ for all values of $\gammaaniso$ under consideration we can exclude that the phase transition is of first order. \\

\begin{figure}[t!]
\begin{center}
\epsfig{file= 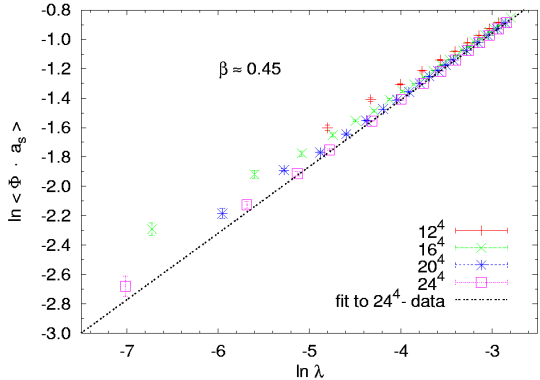, width=8cm}
\caption{Double logarithmic plot of $\langle \Phihat \rangle$ vs. $\lambda$.  }
\label{fig-estimate-beta}
\end{center}
\end{figure}

It is also interesting to numerically determine the exponent $\beta$ 
which is defined by Eq.~\eqref{Def-beta} and which has previously been estimated in the mean-field approximation 
(see Fig.~\ref{fig-MFA-deriv}).  The result of a linear fit to  $\ln \langle \Phihat \rangle$ plotted versus $\ln \lambda$ at $\gammaaniso=1$ is shown in Fig.~\ref{fig-estimate-beta}. We have only fitted the data from the $24^4$- lattice because it provides the best realization of the thermodynamic limit among the available simulation results. The fit yields 
\begin{equation}\label{fit-beta}
\beta = 0.453 \pm 0.002
\end{equation}
where the error estimate gives the statistical error as computed from a 
$\chi^2$- fit again. This result deviates about ten per cent from the 
MFA- estimate quoted in sect. \ref{MFA}. We emphasize that this way to determine $\beta$ is prone to systematic errors which have not been taken into account in the above error estimate. Moreover, we expect that our approach to fit $\beta$ systematically produces too small a value because the curve of $ \langle \Phihat \rangle$ will always be a smooth one when computed on a finite lattice.\\

\begin{figure}[t!]
\begin{center}
\epsfig{file=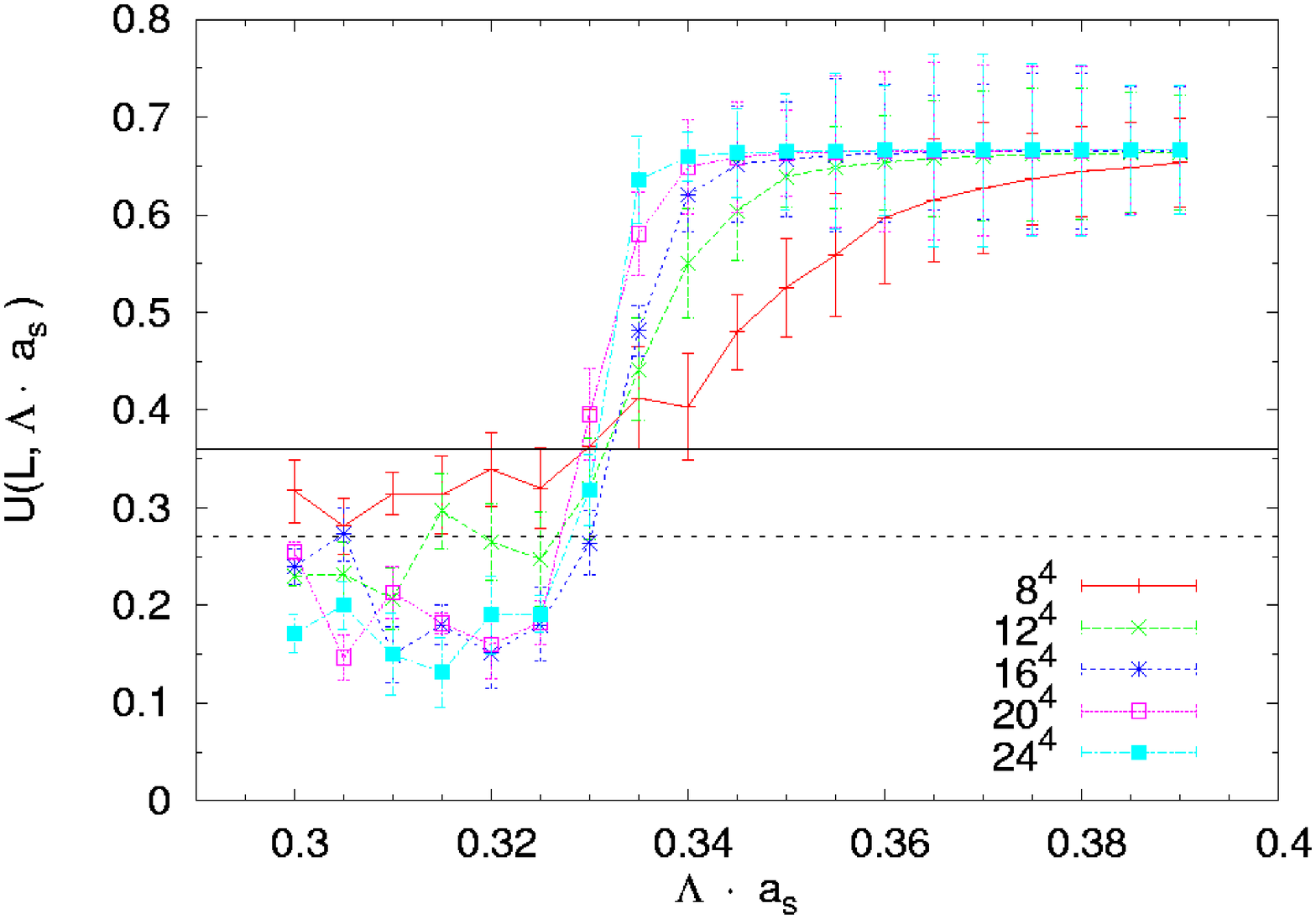, width = 8cm}
\caption{(Color online) The Binder cumulant $U(L, \Lhat)$ plotted vs. $\Lhat$ as computed on lattices of various sizes. The lines connecting the data merely serve to guide the eye. The dashed horizontal line corresponds to the result quoted in \cite{Brezin} while the solid line corresponds to the results of \cite{Lai,Aktekin}.}
\label{fig-Binder-cumulant}
\end{center}
\end{figure}

We conclude this section by presenting the Binder cumulant for the zero temperature case (i. e. $\gammaaniso = 1$) in Fig.~\ref{fig-Binder-cumulant}. The Binder cumulant \cite{Binder} is defined as 
\begin{equation}\label{Def-Binder-cumulant}
U(L , \Lhat) : = 1 - \frac{1}{3} \, \frac{ \langle \, | \Phihat |^4  \, \rangle }{\langle \, | \Phihat |^2 \, \rangle^2 } \; .
\end{equation} 
In Fig.~\ref{fig-Binder-cumulant} the curves intersect at $\Lhat_C \simeq 0.33$, which is in reasonable agreement both with the finite-size scaling result quoted in Tab.~\ref{FSA-results} and the mean-field estimate in Eq.~\eqref{MFA-Lambda_C}. Furthermore, the numerical value of the Binder cumulant at the critical point is characteristic for each universality class. In \cite{Brezin} it was found by analytical means that for the four dimensional Ising model $ \langle s^4 \rangle / \langle s^2 \rangle^2 = 2.188...$ where $s$ is the average magnetization per unit volume. Numerical studies of the four dimensional Ising model\cite{Lai, Aktekin} have yielded $ \langle s^4 \rangle / \langle s^2 \rangle^2 = 1.92 (3) $. The corresponding numerical values for the Binder cumulant $U(L,\Lhat_C)$ as defined in Eq.~\eqref{Def-Binder-cumulant} are 0.27 and 0.36 (independent of $L$, of course), respectively, 
and they are indicated in Fig.~\ref{fig-Binder-cumulant} by the solid and dashed line.  It seems as if our data favors $ \langle s^4 \rangle / \langle s^2 \rangle^2 = 1.92 $. However, the data does not allow for a completely unambiguous discrimination against one of these two alternatives. Nevertheless, we infer that the findings for $U(L, \Lhat_C)$ support the statement that the model at hand at zero temperature falls into the universality class of the 4-d Ising model. \\

\section{Summary and conclusions}\label{Summary}

In this work we have analyzed the theory governed by the Lagrangian
\eqref{Lagrangian} by both 
mean-field and Monte Carlo methods. 
At zero temperature, the mean-field analysis predicts 
the existence of
a continuous phase transition and the universality class of the 
four-dimensional Ising model. Both results 
are in agreement with the Monte Carlo results. 
It seems that the additional symmetry of the potential $V(\Phi) =
V(\Phi^*)$ does 
not have any impact on the critical behavior of the system.

Apart from the phase structure, we have also determined the mass gap  
of the theory both at zero and finite temperature. We found that its 
value is almost independent of temperature. 
The physical anisotropy did not show deviations from the 
bare coupling anisotropy $\gamma$ within the margin of error. 
Finally, we obtained the critical exponents for the transition when 
varying $\Lambda$ also at non-zero temperature from a finite-size
scaling analysis. In the high temperature limit, 
the numerical values seem to  
indicate dimensional reduction to the three dimensional Ising model.
However, more work is necessary to understand the situation 
at high temperature. 

\section*{Acknowledgments}
The numerical computations were carried out on the Fujitsu VPP5000 
of the Rechenzentrum Karlsruhe and on the PC cluster at the 
Institute for Theoretical Physics of the University of Heidelberg. 
Computing support by Werner Wetzel is gratefully acknowledged.\\              

\appendix

\section{Appendix}

\subsection{Some formulae for the MFA}\label{MFA-appendix}

In this appendix we summarize some technical details of the mean-field 
approximation presented in section \ref{MFA}. We use $\gamma=1$. 
It is straightforward to extend the analysis to $\gamma > 1$, however, only for 
$\gamma$ less than about 2 are the results, at least qualitatively, reasonable, 
for larger anisotropy a more involved analysis would be necessary.\\

We start by noting that the functions $F_n(\Lhat, m)$ defined in Eq.~\eqref{Def-F_n} admit the following series expansion in $m$: 
\begin{equation}\label{Expand-F_n}
F_n(\Lhat, m) \; = \; (2d)^n \, \sum_{k=0}^{\infty} \, \frac{1}{k!} (2 d m)^k \, \int_0^{\infty} \, d R \, R^{n +k+1} \, e^{- \tilde{V}(R)} \, \int_0^{2 \pi} \, d \Theta (\cos \Theta)^{ n+k} \, e^{\Lhat^4 \cos ( 2 \Theta ) } \, .
\end{equation}
Note that the terms with odd $n+k$ vanish by virtue of the symmetry properties of the $\Theta$- integral. From this expression we can obtain the derivatives at $m=0$,
to be used in (\ref{MFA-7}), (\ref{expand-m-1}), in terms of
the integrals:
\eab
R_n(\Lhat) &=&\frac{1}{4} \int_0^{\infty} \, d R \, R^{2n + 1} \, e^{- \tilde{V}(R)} 
\; = \; 2 \, \Bigl(  \frac{\Lhat^6 }{ 2 d + \Lhat^2}  
\Bigr)^{\frac{n + 1}{2}} \, K_{n+1}(\Lhat^3 \sqrt{2 d + \Lhat^2 } ) \\
T_n(\Lhat) &=&\frac{1}{4}\,\int_0^{2 \pi} \, d \Theta \, ( \cos^2 \Theta )^n 
\, e^{\Lhat^4 \cos( 2 \Theta ) } \:  = \: \frac{\pi}{2^{n+1}}\,
\sum_{k=0}^n\,\left(\begin{array}{c}n\\ k \end{array}\right)\,
\frac{\partial^k}{\partial u^k} {\rm I}_0(u)|_{u=\Lhat^4} \; .
\eae

The $\omega = \frac{\pi}{2}$ solution can be obtained by reversing
the sign of the $\cos(2\Theta)$ term in \eqref{Expand-F_n}.

\subsection{Implementation of the cluster algorithm}\label{CA-App}

This appendix describes our implementation of a cluster 
algorithm for the model.

We start by rewriting the lattice action Eq.~\eqref{LatticeLagrangian} in terms of the real and imaginary parts $\xhat_n$ and $\yhat_n$ of each field variable $\Phihat_n$, which yields
\begin{equation}\notag
\begin{split}
S[\Phihat] & = \frac{1}{2 \gamma} \, \sum_n \Bigl\{  - \gamma^2 \,  2 \, ( \xhat_n \, \xhat_{n + \hat{\tau}} + \yhat_n \, \yhat_{n + \hat{\tau}} )   - \, 2 \, \sum_j \bigl[ \xhat_n \, \xhat_{n + \hat{j}} + \yhat_n \,  \yhat_{n + \hat{j}}  \bigr]  \\
  & \hspace{1.5cm}  + \,2 \, ( 3 + \gamma^2) \, \vert \Phihat_n \vert^2  + V(\Phihat_n)  \Bigr\} \, .
\end{split}
\end{equation}   
We now identify a set of Ising- like degrees of freedom in the model by defining $s_n := \text{sign}(\xhat_n)= \text{sign }( \text{Re}\,  \Phihat_n  )$ which we will use to set up the cluster algorithm. The action can then be rewritten as $ S = S_{\text{Ising}} + S_{\text{Rest}} $ where $S_{\text{Rest}}$ does not depend on the $\{ s_n \}_n $ and
\begin{equation}\label{csd2}
\begin{split}
S_{\text{Ising}}[ \Phihat]  & = - \: \sum_n \Bigl\{  \gamma | \xhat_n | \, | \xhat_{n + \hat{\tau}} | \, s_n \, s_{n + \hat{\tau}}  +  \sum_j \Bigl( \;  \frac{1}{\gamma}| \xhat_n | \, | \xhat_{n + \hat{j}} | \, s_n \, s_{n + \hat{j}}  \; \Bigr)   \Bigr\}  \\
& = - \: \sum_n \sum_{\mu} J_{n, n + \hat{\mu}} \, s_n \, s_{n + \hat{\mu}}  \\
& = -  \, \frac{1}{2} \, \sum_{< l, n > } J_{l,n} \, s_l \, s_n   \;.
\end{split}
\end{equation} 
Here we have defined a set of site-dependent nearest neighbor couplings as 
\begin{equation}\label{Def-Jmn-CA}
J_{l,n} :=
\begin{cases} \gamma \, | \xhat_l | \, | \xhat_n | &    \text{if l, n are time-like neighbors}  \\
\frac{1}{\gamma} \, | \xhat_l | \, | \xhat_n | &    \text{if l, n  are space-like neighbors}
\end{cases}  \; \; .
\end{equation}
We have chosen to embed Ising spins only in the real part of the field. Due to the shape of the potential we expect the imaginary part to be less affected by critical slowing down, if at all. Moreover, we note that it is not possible to define Ising spins from a projection on a random axis in the complex $\Phihat_n$- plane as it can be done for instance in the case of the XY- model. The reason for this is that the potential does not possess the necessary symmetry; it is in general not possible to flip the projection of $\Phihat_n$ on an arbitrary axis without altering the value of $S_{\text{Rest}}$. \\
It remains to set up a cluster algorithm for the Ising- like degrees of freedom. The dynamics of the spin variables $\{s_n\}_n$ is equivalent to the one of a four dimensional Ising model governed by the Hamiltonian
\begin{equation}\label{IsingE}
H_{\text{Ising}} \equiv S_{\text{Ising}} = - \frac{1}{2} \, \sum_{ < l, n > } J_{l,n} \, s_l \, s_n \, .
\end{equation} 
We then proceed in the standard way \cite{FK,SwendsenWang,Wolff1} by introducing bond variables $d_{l,n}$ for all pairs of nearest neighbors $l$, $n$. From the literature (cf. e. g. \cite{SwendsenWang}) it is known that the probability for a bond to be active has to be chosen as 
\begin{equation}\label{LinkActivProb}
p_{\text{act}}(d_{l,n}) =
\begin{cases} 0 & \text{ if } \: s_l \neq s_n \\
 1 - \exp(- 2 J_{l,n} ) & \text{ if } \: s_l = s_n
\end{cases}
\end{equation}
in order to implement the correct dynamics for the Ising Hamiltonian \eqref{IsingE}. After activating each bond variable with this probability we proceed as described in \cite{Wolff1}: First, we choose a lattice site at random. Then we track the percolation cluster to which this site belongs and flip the real parts $\xhat_n \mapsto - \xhat_n $ of all field variables in the percolation cluster. \\


\begin{thebibliography}{50}

\bibitem{Fisher} M. E. Fisher, Rev. Mod. Phys. 70, 653 (1998)
\bibitem{Ralf2}  R. Hofmann, Int. J. Mod. Phys. A \textbf{20}, 4123
(2005), Erratum-ibid. A {\bf 21}, 6515 (2006).\\ 
R. Hofmann, Mod. Phys. Lett. A {\bf 21}, 999 (2006), Erratum-ibid. A {\bf 21}, 3049 (2006). 
\bibitem{Hofmann2005C} R. Hofmann, hep-th/0508212. 
\bibitem{Hofmann2007} R. Hofmann, Mod. Phys. Lett. A \textbf{22}, 2657 (2007)
\bibitem{Karsch1} J. Engels, F. Karsch, H. Satz, Nucl. Phys. \textbf{B205}, 239 (1982).
\bibitem{Karsch-Nucu} G. Burgers, F. Karsch, A. Nakamura, I. O. Stamatescu, Nucl. Phys. \textbf{B304}, 587 (1988).
\bibitem{stanley} H. E. Stanley, \textit{Introduction to Phase Transitions and Critical Phenomena} , Clarendon Press, Oxford 1971.
\bibitem{Metro} N. Metropolis et. al., J. Chem. Phys. \textbf{21}, 1087 (1953).
\bibitem{Creutz2} M. Creutz et al., \textit{Monte Carlo Computations in Lattice Gauge Theories} , Phys. Rept. \textbf{95}, 201 (1983).
\bibitem{Brower} R. C. Brower, P. Tamayo, Phys. Rev. Lett. \textbf{62}, 1087 (1989).
\bibitem{Sok1} A. D. Sokal, \textit{How to beat critical slowing down}, Nucl. Phys. B (Proc. Suppl.) \textbf{20}, 55 - 67 (1991).
\bibitem{Talapov-Bloethe} A. L. Talapov, H. W. J. Bl\"othe, J. Phys. A, Math. Gen. \textbf{29}, 5727-5734 (1996).
\bibitem{MSchiestl} M. Schiestl, PhD thesis, Universit\"at Heidelberg (1991).
\bibitem{Zinn-Justin} J. C. Le Guillou, J. Zinn-Justin, Phys. Rev. B \textbf{21}, 3976 - 3998 (1980).
\bibitem{Nucu-vL} L. von L\"ohneysen, R. E. Shrock, I. O. Stamatescu, Phys. Lett. B. \textbf{205}, 321-328 (1988).
\bibitem{O-Connor-1} D. O'Connor, C. R. Stephens, J. Phys. A, Math. Gen. \textbf{25}, 101-108, (1992).
\bibitem{O-Connor-2} D. O'Connor, C. R. Stephens,  Nucl. Phys. \textbf{B360}, 297-336 (1991).
\bibitem{Liao-Strickland} S.-B. Liao, M. Strickland, Nucl. Phys. \textbf{B497}, 611-638 (1997).
\bibitem{Fisher-Berker} M. E. Fisher, A. Nihat Berker, Phys. Rev. B \textbf{26}, 2507-2513 (1982).
\bibitem{Cardy} J. L. Cardy, P. Nightingale, Phys. Rev. B \textbf{27}, 4256-4260 (1983).
\bibitem{Binder-Landau} K. Binder, D. P. Landau, Phys. Rev. B \textbf{30}, 1477-1485 (1984).
\bibitem{Binder} K. Binder, Phys. Rev. Lett. \textbf{47}, 693 (1981).
\bibitem{Brezin} E. Br\'ezin, J. Zinn-Justin, Nucl. Phys. \textbf{B257}, 867-893 (1985).
\bibitem{Lai} P.-Y. Lai, K. K. Mon, Phys. Rev. B \textbf{41}, 9257-9263 (1990).
\bibitem{Aktekin} N. Aktekin, J. Stat. Phys. 104, 1397-1406 (2001).
\bibitem{Reinhardt} H. Reinhardt, Nucl. Phys. \textbf{B628}, 133-166 (2002).
\bibitem{Politzer} H. D. Politzer, Phys. Rev. Lett. \textbf{30}, 1346 (1973).
\bibitem{Gross-Wilczek} D. J. Gross, F. Wilczek, Phys. Rev. Lett. \textbf{30}, 1323 (1973).
\bibitem{FK} C. M. Fortuin, P. W. Kasteleyn, Physica \textbf{57}, 536-564 (1972).
\bibitem{SwendsenWang} R.H. Swendsen, J. S. Wang, Phys. Rev. Lett. \textbf{58},  86 (1987).
\bibitem{Wolff1} U. Wolff, Phys. Rev. Lett. \textbf{62}, 361 (1989).
\end{thebibliography}
\end{document}